# Modeling multi-particle complexes in stochastic chemical systems

Muir J. Morrison
*Department of Physics, California Institute of Technology*

Justin B. Kinney
*Simons Center for Quantitative Biology, Cold Spring Harbor Laboratory*[*]

(Dated: March 23, 2016)

Large complexes of classical particles play central roles in biology, in polymer physics, and in other disciplines. However, physics currently lacks mathematical methods for describing such complexes in terms of component particles, interaction energies, and assembly rules. Here we describe a Fock space structure that addresses this need, as well as diagrammatic methods that facilitate the use of this formalism. These methods can dramatically simplify the equations governing both equilibrium and non-equilibrium stochastic chemical systems. A mathematical relationship between the set of all complexes and a list of rules for complex assembly is also identified.

Fock spaces – the vector spaces in which quantum field theories are built – provide a natural way to represent physical systems that have variable particle composition. Multiple Fock space formalisms have been described for modeling stochastic chemical systems (e.g., [1–3]), and these have proven useful in a variety of contexts. In particular, the Fock space methods described by Grassberger and Scheunert [2], for which Peliti introduced a path integral formulation [4], have been widely adopted [5, 6], especially is studies of diffusion-limited processes [7].

When modeling chemical systems that generate large multi-particle complexes, however, these formalisms become problematic. For instance, many interesting chemical systems in biology and in polymer physics are capable of generating vast (or even infinite) numbers of distinct complexes based on a relatively small number of components and interaction rules. Existing formalisms treat each distinct multi-particle complex as its own species of particle. It is often impractical to manually enumerate these complexes, to specify each one's free energy, formation and decay rates, and so on.

The proliferation of complexes and the difficulties it can lead to are well-recognized in the context of molecular systems biology [8–10]. To address this issue computationally, formal grammars [11–15] and accompanying software [16–20] have been developed that enable "rule-based" simulations of biochemical systems. However, a "rule-based mathematics" that allows one to work with such systems analytically has yet to be described.

Here we introduce a mathematical formalism that allows rule-based definitions of both equilibrium and non-equilibrium stochastic chemical systems. The Fock space we describe is most similar to that of Doi [1] and of Park and Park [3], but with a number of key differences. Every particle in this formalism is modeled as occupying one of a large number of internal states. These internal states uniquely identify each particle and are essential for representing multi-particle complexes in terms of their components. Fock space excitations are used to represent not just particles, but also the interactions between particles, the conformation states of particles, and occupied sites on the surfaces of particles. The concept of occlusion, which is essential for any rule-based description of multi-particle complexes, is realized by using a Fock space constructed from hard core bosons [3].

Following [1], we equip our Fock space with an orthonormal basis $\mathcal{S}$ of "pure" physical states. The vector $|\psi\rangle$ that describes a system of interest is given by

$$|\psi\rangle = \sum_{s \in \mathcal{S}} p_s |s\rangle, \qquad (1)$$

where $p_s$ is the probability that the system, when observed, will be found in state $s$. Every measurable quantity $Q$ is represented by a corresponding operator $\mathbb{Q}$ that is diagonal in $\mathcal{S}$, and the expectation value for this quantity is given in terms of $|\psi\rangle$ by $\langle Q \rangle = \langle \text{sum}| \mathbb{Q} |\psi\rangle$, where

$$|\text{sum}\rangle \equiv \sum_{s \in \mathcal{S}} |s\rangle \qquad (2)$$

is referred to as the "sum vector." In thermal equilibrium, $|\psi\rangle$ is uniquely determined by $|\text{sum}\rangle$ and a "Hamiltonian" operator $\mathbb{H}$ that assigns a free energy to each pure state:

$$|\psi\rangle = \frac{e^{-\mathbb{H}/kT}}{Z} |\text{sum}\rangle, \qquad (3)$$

where the partition function $Z$ is given by

$$Z \equiv \langle \text{sum}| e^{-\mathbb{H}/kT} |\text{sum}\rangle. \qquad (4)$$

Outside of thermal equilibrium, time evolution of the system is governed by a master equation

$$\frac{d}{dt} |\psi\rangle = \left( \mathbb{R} - \mathring{\mathbb{R}} \right) |\psi\rangle, \qquad (5)$$

where $\mathbb{R}$ is a "rate matrix" and $\mathring{\mathbb{R}}$ is a closely related "depletion matrix." In terms of the transition rates $R_{s \to t}$ from any pure state $s$ to any other pure state $t$, these operators are given by

$$\mathbb{R} = \sum_{s,t \in \mathcal{S}} R_{s \to t} |t\rangle \langle s|, \quad \mathring{\mathbb{R}} = \sum_{s,t \in \mathcal{S}} R_{s \to t} |s\rangle \langle s|. \qquad (6)$$

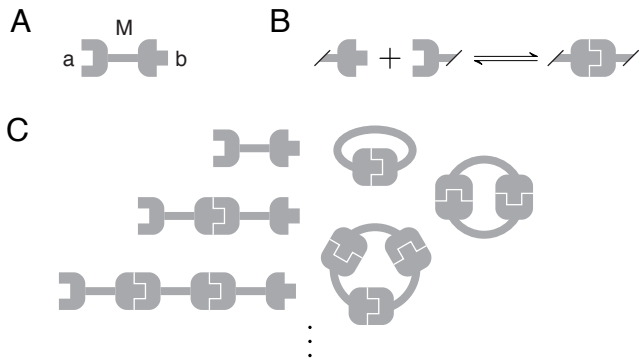

FIG. 1. The 0-dimensional polymer model. (A) Each monomer $M$ has two binding sites, $a$ and $b$. (B) Any unoccupied $a$ site is allowed to bind to any unoccupied $b$ site. (C) This system can generate an infinite number of both linear and circular complexes.

To represent a 0-dimensional gas of monomeric particles (called $M$), we use a Fock space containing a field that has $N$ modes $\{M_i\}_{i=1}^N$. Each mode $M_i$ represents a single internal state of the particle $M$ and behaves as a hard core boson [3]. Specifically, there is a raising operator $\hat{M}_i$ and a lowering operator $\check{M}_i \equiv \hat{M}_i^\dagger$ that are nilpotent ($\hat{M}_i^2 = \check{M}_i^2 = 0$) and that satisfy the anticommutation relation $\{\check{M}_i, \hat{M}_i\} \equiv 1$. For each mode we further define a "presence" operator $\bar{M}_i \equiv \hat{M}_i \check{M}_i$ and an "absence" operator $\tilde{M}_i \equiv \check{M}_i \hat{M}_i$. All operators for different modes commute, and the vacuum state $|0\rangle$ is defined to be annihilated by every annihilation operator $\check{M}_i$. All calculations are performed in the $N \to \infty$ limit.

The sum vector is given by the sum of the vacuum state, all 1-particle states, all 2-particle states, etc.. Written in terms of $|0\rangle$ and the raising operators $\hat{M}_i$, this is

$$|\text{sum}\rangle = |0\rangle + \sum_i \hat{M}_i |0\rangle + \frac{1}{2!} \sum_{i,j} \hat{M}_i \hat{M}_j |0\rangle + \cdots . \quad (7)$$

Note that each term in this series is multiplied by an inverse factorial coefficient that compensates for overcounting due to permutation symmetry of the summands. We can therefore write

$$|\text{sum}\rangle = e^{\mathbb{G}} |0\rangle , \quad (8)$$

where $\mathbb{G} \equiv \sum_i \hat{M}_i$. Eq. (8) in fact holds for any chemical system so long as $\mathbb{G}$ equals the sum of creation operators for all internal states of all possible complexes. This fact was noted by Doi [1], but the operator $\mathbb{G}$ does not yet have an accepted name. Here we call $\mathbb{G}$ the "gallery".

In thermal equilibrium, $|\psi\rangle$ is fully specified by a chemical potential $\mu$. The Hamiltonian corresponding to this chemical potential is

$$\mathbb{H} = -\mu' \sum_{i=1}^N \bar{M}_i, \quad (9)$$

where $\mu' = \mu - kT \ln N$ is the chemical potential appropriately adjusted for the number of internal states of the monomer. Expanding Eq. (3) and Eq. (4) term-by-term, one readily verifies that this gallery $\mathbb{G}$ and Hamiltonian $\mathbb{H}$ properly define a zero-dimensional gas of monomeric particles in the grand canonical ensemble.

Now suppose the particles $M$ are capable of forming directed homopolymer chains, as depicted in Fig. 1. The gallery for this system is given by

$$\mathbb{G} = \mathbb{G}_{1l} + \mathbb{G}_{1c} + \mathbb{G}_{2l} + \mathbb{G}_{2c} + \mathbb{G}_{3l} + \mathbb{G}_{3c} + \cdots \quad (10)$$

where $\mathbb{G}_{ml}$ and $\mathbb{G}_{mc}$ respectively denote the sum of creation operators for linear and circular polymer chains having $m$ subunits. If each polymer chain were represented by a separate field, the corresponding Hamiltonian would, by analogy to Eq. (9), require an infinite number of terms (one for each complex), each multiplied by its own chemical potential. This proliferation of terms and parameters is inconsistent with our expectation that the Hamiltonian should describe the essential energetic contributions to a system, since the polymer system in question is fully specified by just two parameters: the chemical potential $\mu$ of monomeric particles and the interaction energy $\epsilon$ between pairs of bound particles.

This problem is remedied by representing each multiparticle complex in terms of its components and interactions. To do this we introduce the "site fields" $a$ and $b$, which have $N$ modes each ($\{a_i\}_{i=1}^N$ and $\{b_i\}_{i=1}^N$), as well as an "interaction field" $I$ that has $N^2$ modes $\{I_{ij}\}_{i,j=1}^N$. Like $M$, the fields $a$, $b$, and $I$ behave as hard core bosons. Table I shows each term of the gallery in Eq. (10) expressed in terms of these fields. The creation operator for the linear dimer, for example, is given by

$$\hat{D}_{ij} \equiv \hat{M}_i \hat{a}_i \hat{I}_{ij} \hat{b}_j \hat{M}_j. \quad (11)$$

Here, $\hat{M}_i$ and $\hat{M}_j$ create the two component particles of the dimer, $\hat{I}_{ij}$ registers an interaction between these particles, and $\hat{a}_i$ and $\hat{b}_j$ mark the resulting occupied sites. Eq. (11) defines a "composite field" $D$ which has $N^2$ modes that also behave as hard core bosons.

We now describe diagrammatic methods that aid the use of this composite Fock space formalism. Internal indices are represented by dots, pairs of indices by edges connecting two dots, and sets of three or more indices by dots contained within bubbles. Symbols written next to dots, lines, and bubbles indicate operators that have those corresponding internal indices. Conversely, a dot written next to any symbol indicates whatever internal indices that symbol might possess. All operators depicted in a diagram must commute with each other. For example, Eq. (11) can be written as

$$\underset{\hat{D}}{\mathbf{o}} \equiv \underset{\hat{M} \quad \hat{I} \quad \hat{M}}{\overset{\hat{a} \qquad \hat{b}}{\mathbf{o}\!\!\rightarrow\!\!\mathbf{o}}} . \quad (12)$$

TABLE I. Formulas and diagrams for each term of the polymer gallery shown in Eq. (10). For $\mathbb{G}_{1c}$, $\mathbb{G}_{2c}$, and $\mathbb{G}_{3c}$, field names are hidden for conciseness. Note that the coefficients in $\mathbb{G}_{2c}$ and $\mathbb{G}_{3c}$ compensate for over-counting that results from the cyclic permutation symmetry of the summands.

| Term | Formula | Diagram |
|---|---|---|
| $\mathbb{G}_{1l}$ | $\sum_i \hat{M}_i$ | ● $\hat{M}$ |
| $\mathbb{G}_{2l}$ | $\sum_{i,j} \hat{M}_i \hat{a}_i \hat{I}_{ij} \hat{b}_j \hat{M}_j$ | ●→● |
| $\mathbb{G}_{3l}$ | $\sum_{i,j,k} \hat{M}_i \hat{a}_i \hat{I}_{ij} \hat{b}_j \hat{M}_j \hat{a}_j \hat{I}_{jk} \hat{b}_k \hat{M}_k$ | ●→●→● |
| $\mathbb{G}_{1c}$ | $\sum_i \hat{M}_i \hat{a}_i \hat{I}_{ii} \hat{b}_i$ | (loop) |
| $\mathbb{G}_{2c}$ | $\frac{1}{2} \sum_{i,j} \hat{M}_i \hat{a}_i \hat{I}_{ij} \hat{b}_j \hat{M}_j \hat{a}_j \hat{I}_{ji} \hat{b}_i$ | (2-cycle) |
| $\mathbb{G}_{3c}$ | $\frac{1}{3} \sum_{i,j,k} \hat{M}_i \hat{a}_i \hat{I}_{ij} \hat{b}_j \hat{M}_j \hat{a}_j \hat{I}_{jk} \hat{b}_k \hat{M}_k \hat{a}_k \hat{I}_{ki} \hat{b}_i$ | (triangle) |

Sums over internal indices are represented by filling in the appropriate dots (e.g., see Table I). Within such sums, each distinguishable state is counted exactly once. As with Feynman diagrams, this often leads to symmetry factors in the corresponding formulas.

Using this composite Fock space dramatically simplifies the Hamiltonian of the 0D polymer system. Instead of expressing $\mathbb{H}$ as a sum of an infinite number of terms, each multiplied by its own chemical potential, $\mathbb{H}$ can be expressed as a sum of only two terms, one for $\mu$ and one for $\epsilon$:

$$\mathbb{H} = -\mu' \sum_i \bar{M}_i + \epsilon \sum_{i,j} \bar{I}_{ij} = -\mu' \left( \begin{array}{c} \bullet \\ \bar{M} \end{array} \right) + \epsilon \left( \begin{array}{c} \bullet \\ \bar{I} \end{array} \right). \quad (13)$$

This Hamiltonian evaluates the energy of each complex in a simple and intuitive way: the first term contributes a free energy of $-\mu$ for each component particle, while the second term contributes a free energy of $\epsilon$ for every two-particle interaction. Note the use of a single dot next to $I$; since the internal state of $I$ is indexed by the pair $ij$, this dot indicates summation over both indices.

This composite Fock space can also dramatically simplify master equations. Suppose that each potential $a{:}b$ interaction forms at a rate $r_+$, while each realized interaction decays at a rate $r_-$. The corresponding rate matrix is given by

$$\mathbb{R} = r_+ \sum_{i,j} \hat{a}_i \hat{b}_j \hat{I}_{ij} \bar{M}_i \bar{M}_j + r_- \sum_{i,j} \check{a}_i \check{b}_j \check{I}_{ij} \bar{M}_i \bar{M}_j \quad (14)$$

$$= r_+ \left( \begin{array}{cc} \hat{a} & \hat{b} \\ \bullet \!\!\rightarrow\!\! \bullet \\ \bar{M} \ \hat{I} \ \bar{M} \end{array} \right) + r_- \left( \begin{array}{cc} \check{a} & \check{b} \\ \bullet \!\!\rightarrow\!\! \bullet \\ \bar{M} \ \check{I} \ \bar{M} \end{array} \right). \quad (15)$$

The first term links two $M$ particles together and registers the appropriate $a$ and $b$ sites as occupied. The second term destroys preexisting interactions, in the process freeing up sites $a$ and $b$. Note that the two diagrams in Eq. (15) are Hermitian conjugates of one another. The corresponding depletion matrix follows from $\mathbb{R}$ by replacing creation operators with absence operators and annihilation operators with presence operators:

$$\dot{\mathbb{R}} = r_+ \left( \begin{array}{cc} \tilde{a} & \tilde{b} \\ \bullet \!\!\rightarrow\!\! \bullet \\ \bar{M} \ \tilde{I} \ \bar{M} \end{array} \right) + r_- \left( \begin{array}{cc} \bar{a} & \bar{b} \\ \bullet \!\!\rightarrow\!\! \bullet \\ \bar{M} \ \bar{I} \ \bar{M} \end{array} \right). \quad (16)$$

This method for transforming rate matrices into depletion matrices is fully general. The ease with which this master equation is defined stands in stark contrast to the difficulty of manually specifying correct transition rates between every one of the infinite linear and circular polymer species illustrated in Fig. 1C.

We now address the problem of specifying $|\text{sum}\rangle$. Eq. (8) greatly simplifies this task, but manually defining the gallery $\mathbb{G}$ by listing every possible complex can still be cumbersome. The composite Fock space enables an alternative approach: $|\text{sum}\rangle$ can be defined using a "factory" – an ordered list of operators that specify rules for assembling complexes. Writing the factory as $\mathcal{F} = (\mathbb{F}_1, \cdots, \mathbb{F}_K)$, the sum vector is given by

$$|\text{sum}\rangle = e^{\mathbb{F}_K} \cdots e^{\mathbb{F}_1} |0\rangle. \quad (17)$$

The order of operators in the factory is important because they do not commute. Indeed, the fact that they do not commute is what generates nontrivial complexes.

The 0D polymer system is readily defined by a two-operator factory $\mathcal{F} = (\mathbb{F}_1, \mathbb{F}_2)$, where $\mathbb{F}_1$ (same as $\mathbb{G}_{1l}$) creates isolated particles and $\mathbb{F}_2$ (which appears in $\mathbb{R}$) binds two preexisting particles together:

$$\mathbb{F}_1 = \begin{array}{c} \bullet \\ \hat{M} \end{array}, \quad \mathbb{F}_2 = \begin{array}{c} \hat{a} \quad \hat{b} \\ \bullet \!\!\rightarrow\!\! \bullet \\ \bar{M} \ \hat{I} \ \bar{M} \end{array}. \quad (18)$$

To verify $e^{\mathbb{F}_2} e^{\mathbb{F}_1} |0\rangle = e^{\mathbb{G}} |0\rangle$, one can Taylor expand the left-hand-side and examine it term by term. For instance, applying the fifth-order term of $e^{\mathbb{F}_1}$ and the second-order term of $e^{\mathbb{F}_2}$ to $|0\rangle$ yields all pure states constructed from five particles and two interactions:

$$\frac{1}{2!} \left( \begin{array}{c} \hat{a} \ \hat{b} \\ \bullet\!\!\rightarrow\!\!\bullet \\ \bar{M} \hat{I} \bar{M} \end{array} \right)^2 \frac{1}{5!} \left( \begin{array}{c} \bullet \\ \hat{M} \end{array} \right)^5 |0\rangle = \Big\{ (\bullet) \frac{1}{2!} (\bullet\!\!\rightarrow\!\!\bullet)^2 + \quad (19)$$

$$\frac{1}{2!} (\bullet)^2 (\bullet\!\!\rightarrow\!\!\bullet\!\!\rightarrow\!\!\bullet) + \frac{1}{2!} (\bullet)^2 (\varhexagon) (\bullet\!\!\rightarrow\!\!\bullet) +$$

$$\frac{1}{3!} (\bullet)^3 \frac{1}{2!} (\varhexagon)^2 + \frac{1}{3!} (\bullet)^3 (\bigcirc) \Big\} |0\rangle.$$

As in Table I, field names are hidden on the right hand side of Eq. (19) for clarity. It is straight-forward, if tedious, to verify Eq. (19) using the commutation relation $[\bar{M}_i, \hat{M}_j] = \delta_{ij} \hat{M}_i$. In doing so one sees that Eq. (17) successfully accounts for subtle combinatorial effects, such as the factor of $\frac{1}{2}$ in $\mathbb{G}_{2c}$ that arises due to cyclic permutation symmetry. One also sees how the hard core boson nature of site fields $a$ and $b$ prevents unphysical complexes from forming. We note that using diagrams greatly eases this computation, allowing the right-hand-side of Eq. (19) to be computed by inspection.

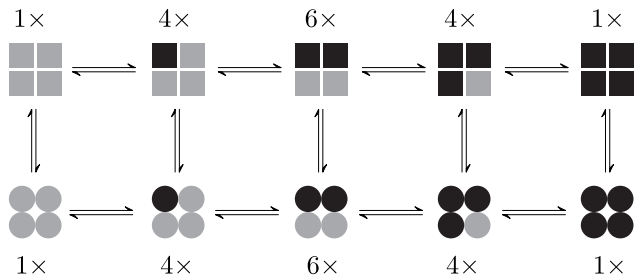

FIG. 2. A standard depiction of the MWC model. Squares represent hemoglobin monomers in the tense conformation, circles represent the relaxed conformation. Subunits bound by oxygen are indicated in black; those not bound are shown in gray. The number of distinct tetramer-oxygen complexes corresponding to each diagram is indicated.

TABLE II. Factory for the MWC model.

| Term | Formula | Diagram |
|---|---|---|
| $\mathbb{F}_1$ | $\frac{1}{4!} \sum_{i,j,k,l} (\hat{H}_4)_{ijkl} \hat{H}_i \hat{H}_j \hat{H}_k \hat{H}_l$ | $\hat{H}_4$ / $(\hat{H} \bullet \bullet \hat{H} / \hat{H} \bullet \bullet \hat{H})$ |
| $\mathbb{F}_2$ | $\frac{1}{4!} \sum_{i,j,k,l} (\bar{H}_4)_{ijkl} (\hat{T}_4)_{ijkl} \hat{T}_i \hat{T}_j \hat{T}_k \hat{T}_l$ | $\hat{T}_4 \ \bar{H}_4$ / $(\hat{T} \bullet \bullet \hat{T} / \hat{T} \bullet \bullet \hat{T})$ |
| $\mathbb{F}_3$ | $\sum_i \bar{H}_i \hat{o}_i$ | $\bar{H} \bullet \hat{o}$ |

To illustrate the utility of these composite Fock spaces for describing more complex biochemical systems, we turn to the classic MWC model [21] for the cooperative binding of oxygen by hemoglobin. In this model, hemoglobin proteins are allowed to be in two conformations, relaxed or tense. These proteins exist only as tetramers in solution, however, and all proteins in the same tetramer must be in the same conformation. The cooperative binding of oxygen results from tense proteins being energetically favored in the absence of oxygen, whereas proteins in the relaxed conformation bind oxygen more tightly.

The molecular complexes of the MWC model can be defined by a three-term factory $\mathcal{F} = (\mathbb{F}_1, \mathbb{F}_2, \mathbb{F}_3)$, the operators of which are shown in Table II. $\mathbb{F}_1$ creates four hemoglobin monomers, represented by the field $H$, which are in the relaxed state by default. It also links them together into a tetramer, represented by the field $H_4$. $\mathbb{F}_2$ transforms a hemoglobin tetramer and its individual subunits from the relaxed to the tense conformation; tense conformations are represented by the fields $T$ and $T_4$ for the individual particles and for the tetramer, respectively. $\mathbb{F}_3$ binds oxygen to a preexisting hemoglobin monomer. Oxygen is not modeled explicitly, but rather its occupancy is indicated only by the site field $o$. The corresponding Hamiltonian is

$$\mathbb{H} = -\mu'_{H_4} (\ \bar{H}_4 \bullet\ ) - kT \ln \alpha (\bar{o} \bullet) \qquad (20)$$
$$- kT \ln L (\bar{T}_4 \bullet) - kT \ln c (\bar{o} \bullet \bar{T}).$$

Here, $\mu'_{H_4}$ is the adjusted chemical potential of hemoglobin tetramers, while $\alpha$, $L$, and $c$ are the original parameters described in [21]: $\alpha$ governs the probability of a relaxed monomer binding oxygen, $L$ governs the relative proportion of tense versus relaxed tetramers in the absence of oxygen, and $c$ quantifies the change in affinity for oxygen of tense versus relaxed monomers. Cooperativity obtains when $L > 1$ and $c < 1$.

This depiction of the MWC model is far more concise and rigorous than the standard representation (e.g., [22–25]). Typically, an illustration similar to Fig. 2 is supplemented with text explaining how to interpret it mathematically. By contrast, the factory in Table II and the Hamiltonian in Eq. (20) provide a fully rigorous mathematical specification of the MWC model; additional text is needed only to provide a biochemical interpretation.

The formalism described here provides a powerful way to concisely and rigorously define mathematical models of stochastic chemical systems that generate large multi-particle complexes. This approach bridges the gap between the mathematical methods used to describe simple stochastic chemical systems and the rule-based approaches that have been developed for computationally simulating more complex systems. One result of this formalism is the identification of a relationship between the set of possible complexes and a list of assembly rules (Eq. (17)). The concepts of space and orientation, which have been ignored thus far, are readily incorporated the way they are in other formalisms (e.g., [1, 3]), i.e., by introducing additional indices. We anticipate that the mathematical and diagrammatic methods described here will prove particularly useful for studying complex biochemical systems, both analytically and computationally.

We thank Rob Phillips for inspiring our work on this problem, as well as Jané Kondev, Ilya Nemenman, and Bruce Stillman for helpful discussions. The work of MJM was supported by NSF Graduate Fellowship DGE-1144469. JBK acknowledges support from the Simons Center for Quantitative Biology at Cold Spring Harbor Laboratory.